\documentclass[a4paper]{jpconf}
\usepackage{graphicx}
\usepackage{cite}

\begin{document}
\title{Astrometric tests of General Relativity in the Solar System: mathematical and computational scenarios}

\author{A Vecchiato$^1$, M Gai$^1$, M G Lattanzi$^1$, M Crosta$^1$, U Becciani$^2$ and S Bertone$^{1,3}$}

\address{$^1$INAF - Astrophysical Observatory of Torino, via Osservatorio 20, 10025 Pino Torinese (TO), Italy}
\address{$^2$INAF - Astrophysical Observatory of Catania, via S. Sofia 78, 95123 Catania (CT), Italy}
\address{$^3$Observatory of Paris, 61 Avenue de l'Observatoire, 75014 Paris, France}

\ead{vecchiato@oato.inaf.it}

\begin{abstract}
We review the mathematical models available for relativistic astrometry,
discussing the different approaches and their accuracies in the context of
the modern experiments from space like Gaia and GAME, and we show how these
models can be applied to the real world, and their consequences from the
mathematical and numerical point of view, with specific reference to the case
of Gaia, whose launch is due before the end of the year.
\end{abstract}

\section{Introduction}
Tests of gravity theories within the Solar System are usually analyzed in the
framework of the so-called \emph{Parametrized Post-Newtonian framework} which
enables the comparison of several theories through the estimation of the value
of a limited number of parameters. Among these parameters, $\gamma$ and $\beta$
are of particular importance for astrometry since they are connected with the
classical astrometric phenomena of the light deflection and of the excess of
perihelion precession in the orbits of massive objects. The same parameters
are of capital importance in fundamental physics, for the problem of characterizing
the best gravity theory, and for the Dark Energy/Dark Matter \cite{2002PhRvL..89h1601D,2006MPLA...21.2291C}.
Moreover, precise astrometric measurements are also important in other tests
of fundamental physics \cite{2005A&A...431..385I,2010ExA....28..209T,2011Ap&SS.331..351I}
since, e.g., they have the potentiality to improve on the ephemeris of the
Solar System bodies. These are the reasons why Solar System astrometric
experiments like Gaia and other projects presently under study \cite{2001AAp...369..339P,2004AN....325..267T,2012ExA....34..165G}
have received a particular attention from the community of fundamental physicists
and cosmologists.

Such kind of experiments, however, call for a reliable model applicable to the
involved astrometric measurements which has not only to include a correct
relativistic treatment of the propagation of light, but a relativistic treatment
of the observer and of the measures as well.

At the same time, the large amount of data to be processed, and the complexity
of the problem to be solved, call for the use of High-Performance Computing (HPC)
environments in the data reduction.

\section{Astrometric models}
The development of an astrometric model based on a relativistic framework can be
dated back to at least 25 years ago \cite{1989racm.book.....S,1991ercm.book.....B}.
In their seminal work of 1992 Klioner and Kopeikin \cite{1992AJ....104..897K}
described a relativistic astrometric model accurate to the $\mu$as level foreseen
for the next generation astrometric missions. This model is built in the framework
of the post-Newtonian (pN) approximation of GR, where the finite dimensions and
angular momentum of the bodies of the Solar System are included and linked to the
motion of the observer in order to consider the effects of parallax, aberration,
and proper motion, and the light path is solved using a matching technique that
links the perturbed internal solution inside the near zone of the Solar
System with the assumed flat external one. The light trajectory is solved in a
perturbative way, as a straight line plus integrals containing the
perturbations which represent, i.e., the effects of the aberrational terms, of
the light deflection, etc. An extension of this model accurate to $1~\mu$as called
GREM (Gaia RElativistic Model) was published in 2003 \cite{2003AJ....125.1580K}.
This has been adopted as one of the two model for the Gaia data reduction, and it
is formulated according to the PPN (Parametrized Post-Newtonian) formalism in order
to include the estimation of the $\gamma$ parameter.

A similar approach was followed by Kopeikin, Sch\"afer, and Mashhoon \cite{1999PhRvD..60l4002K,2002PhRvD..65f4025K}
in the post-Minkowskian approximation. In this case, however, the authors used a
Li\'enard-Wiechert representation of the metric tensor to describe a retarded type
solution of the gravitational field equations and to avoid the use of matching
techniques to solve the geodesic equations.

RAMOD (Relativistic Astrometric MODel) is another family of models, whose development
started in 1995. In this approach the definition of the observable according to the
theory of measure \cite{2010cmcs.book.....D} and the immediate application to the
problem of the astrometric sphere reconstruction was privileged. As a consequence, it
started as a simplified model \cite{1998AAp...332.1133D} based on a plain Schwarzschild
metric. Further enhancement brought to the first realistic estimation of the performances
of Gaia for the determination of the PPN $\gamma$ parameter \cite{2003AAp...399..337V},
and to the development of a fully accurate N-Body model of the light propagation and of
an observer suitable for application to space missions \cite{2003CQGra..20.4695B,2006ApJ...653.1552D}.
Since the so-called RAMOD3 \cite{2004ApJ...607..580D}, this model was built on a
complete pM background, and the light propagation was described with the equation of
motion of measurable quantities varying all along the geodesic connecting the starting
point to the observer. This approach brought to a specific form of the geodesic equations
as a set of coupled nonlinear differential equations which could be solved only by
numerical integration. This represented a problem for an extensive application of this
model to practical astrometric problems, which has been solved only recently for RAMOD3
by Crosta \cite{2011CQGra..28w5013C} who applied a re-parametrization of these equations
of motion to demonstrate their equivalence to the model in \cite{2002PhRvD..65f4025K},
thus opening the road to an analytical solution of RAMOD3 \cite{2013arXiv1305.4824C} and
to its full application to astrometry problems.

Finally, another class of models based on the Time Transfer Function (TTF) technique,
has been developed since 2004 \cite{2004CQGra..21.4463L}. The TTF formalism stands as
a development of the Synge World Function \cite{1964rgt..book.....S} which, contrary
to all the method described so far, is an integral approach based on the principle of
least action. In this models one does not solve the system of differential equations
of the geodesic equations, and thus does not retrieve the solution of the equations of
motion of the photons, but it concentrates on obtaining some essential information
about the propagation of these particles between two points at finite distance; the
coordinate time of flight, the direction triple of the light ray at either the point of
emission ($A$) and of reception ($B$), and the ratio ${\cal K}\equiv\left(k_0\right)_B/
\left(k_0\right)_A$ of the temporal components of the tangent four-covector $k_\alpha$,
which is related to the frequency shift of a signal between two points.

\section{Analytical and numerical comparison}
All these models are conceived to be used at least at the $\mu$as level, suitable for
the accuracy foreseen by future astrometric experiments like Gaia and GAME \cite{2001AAp...369..339P,2012ExA....34..165G}.
Nonetheless it has to be considered that, because of the unprecedented level of
accuracy which is going to be reached, both the astrometric models and the data
processing software will be applied for the first time to a real case. Moreover, in
the case of Gaia this problem is even more delicate since here the satellite is
self-calibrating and will perform {\em absolute measurements} which is equivalent to
the definition of a unit of measure. These are some of the reasons why extensive
analytical and numerical comparisons among the different models are being conducted.

From the theoretical and analytical point of view, a first comparison was conducted
in \cite{2010A&A...509A..37C} showing that GREM and RAMOD3 have an equivalent treatment
of the aberration. Later the equivalence of RAMOD3 and the model in \cite{1999PhRvD..60l4002K,2002PhRvD..65f4025K}
at the level of the (differential) equations of motion has then been shown
in \cite{2011CQGra..28w5013C}, while the explicit formulae for the light
deflection and the flight time of GREM, RAMOD3, and TTF was compared in \cite{2013Submitted}
where it is demonstrated the equivalence of TTF and GREM at 1PN in a time-dependent
gravitational field and that of TTF and RAMOD in the static case.

Numerical comparisons between the GREM and the pM models showed that they give the
same results at the sub-$\mu$as level \cite{2003A&A...410.1063K}. On the other side,
GREM has been compared with a low-accuracy ($(v/c)^2$) version of RAMOD proving that
even a relatively unsophisticated modeling of the planetary contributions can take
into account of the light deflection up to the $\mu$as level almost everywhere in the
sky. This means that, in principle, some experiments like the reconstruction of
the global astrometric sphere of Gaia could initially be done by $(v/c)^2$ models.

Both the analytical and the numerical comparison, however, showed that the correct
computation of the retarded distance of the (moving) perturbing bodies is fundamental
to achieve the required accuracy.



\section{Data reduction algorithms: the case of Gaia}
The reduction of the data coming from astrometric missions bring to the attention
of the scientific community another kind of new problems, i.e. those connected
to the need of reducing a huge amount of astrometric data in ways that were
never experienced before. A significant example is given by the problem of the
reconstruction of the global astrometric sphere in the Gaia mission.

From a mathematical point of view, the satellite observations translate into
a large number of equations, linearized with respect to the unknown parameters
around known initial values, which constitute an overdetermined and sparse
equations system that is solved in the least-squares sense to obtain the
astrometric catalog with its errors. In the Gaia mission these tasks are done
by the Astrometric Global Iterative Solution (AGIS) but the international
consortium which is in charge of the reduction of the Gaia data decided to
produce also an independent sphere reconstruction named AVU-GSR.

This was motivated by the absolute character of these results, and by uniqueness
of the problem which comes from several factors, the main being represented by
the dimensions of the system which are of the order of $10^{10}\times10^8$.
A brute-force solution of such system would require about $10^{27}$ FLOPs, a
requirement which cannot be decreased at acceptable levels even considering that
the sparsity rate of the reduced normal matrix is of the order of $10^{-6}$. It is
therefore necessary to resort to iterative algorithms.

AGIS uses additional hypotheses on the correlations among the unknowns which
are reflected on the convergence properties of the system and permit a separate
adjustment of the astrometric, attitude, instrument calibration, and global
parameters, allowing the use of an embarrassingly parallel algorithm \cite{2012A&A...538A..78L}.
The starting hypotheses, however, can hardly be proved rigorously, and have only
be verified ``a posteriori'' by comparing the results with simulated true values,
a situation which cannot hold in the operational phase with real data. Moreover,
this method by definition prevents the estimation of the correlations between
the different types of unknown parameters, which constitute the other unique
characteristic of this problem. These considerations about the AGIS module lead
to the solution followed by AVU-GSR, which uses a modified LSQR algorithm \cite{1982ATMS....8...43P})
to solve the system of equations which, however, cannot be solved without
resorting to HPC parallel programming techniques as explained in \cite{10.1109/WETICE.2012.66}.


\section{Conclusions}
The increasing precision in the modern astrometric measurements from space makes
high-accuracy tests of the DM/DE vs. Gravity theory debate a target accessible to
future space-born astrometric missions. To this aim, viable relativistic astrometric
models are needed, and three classes of models have been developed during the last
two decades. Work is still on-going to cross-check their mutual compatibility at
their full extent, but what has been done so far demonstrated that they are equivalent
at least at the level of accuracy required for the Gaia measurements. At the same time
these missions put new challenges to the efforts of data reduction. We have briefly
shown how the problem was faced in Gaia, in the limited contest of the reconstruction
of the global astrometric sphere, where an additional constraint is put by the absolute
character of its main product.

\ack
This work has been partially funded by ASI under contract to INAF I/058/10/0
(Gaia Mission - The Italian Participation to DPAC).

\section*{References}
\providecommand{\newblock}{}

\end{document}